\documentclass[twocolumn,amsmath,amssymb,prl]{revtex4}

\usepackage{graphicx}
\usepackage{dcolumn}
\usepackage{bm}
\usepackage{color}
\usepackage{notes2bib}

\newcommand{\rmi}{{\rm i}}

\begin{document}

\title{Optically detected spin-mechanical resonance in silicon carbide membranes}

\author{A.\,V.\,Poshakinskiy$^{1}$}
\author{G.\,V.\,Astakhov$^{2,1}$}

\affiliation{$^1$Ioffe Institute, 194021 St.~Petersburg, Russia  \\ 
$^2$Helmholtz-Zentrum Dresden-Rossendorf, Institute of Ion Beam Physics and Materials Research, 01328 Dresden, Germany }

\begin{abstract}  
Hybrid spin-mechanical systems are a promising platform for future quantum technologies. Usually they require application of additional microwave fields to project integer spin to a readable state. We develop a theory of optically detected spin-mechanical  resonance  associated with half-integer spin defects in silicon carbide (SiC) membranes. It occurs when a spin resonance frequency matches a resonance frequency of a mechanical mode, resulting in a shortening of the spin relaxation time through resonantly enhanced spin-phonon coupling. The effect can be detected as an abrupt reduction of the photoluminescence intensity under optical pumping without application of microwave fields. We propose all-optical protocols based on such spin-mechanical  resonance to detect external magnetic fields and mass  with ultra-high sensitivity. We also discuss room-temperature nonlinear effects under strong optical pumping, including spin-mediated cooling and heating of mechanical modes. Our approach suggests a new concept for quantum sensing using spin-optomechanics. 
\end{abstract}

\date{\today}

\maketitle
 
 \section{Introduction}
 
Cavity optomechanics is an emerging research field exploring the interaction between electromagnetic radiation and mechanical resonators \cite{Aspelmeyer:2014ce}. The motivation for the research originates from exciting fundamental physics as well as various technological applications, including high-resolution accelerometers \cite{Krause:2012cf} and quantum transducers \cite{Bochmann:2013di}. The interaction between light and nanomechanical modes  can also be mediated by electron spin qubits \cite{Maze:2011gw, Doherty:2011bq, Udvarhelyi:2018bx}. Indeed, such a hybrid spin-mechanical quantum system have been realized using the NV center in diamond  \cite{Arcizet:2011cg, Kolkowitz:2012iw, Ovartchaiyapong:2014gv, MacQuarrie:2015ie, Barfuss:2015hv, Golter:2016cd, MacQuarrie:2017hl, Barson:2017ba}. Silicon carbide (SiC) is a natural platform for spin optomechanics, as it is used as a material for ultra-sensitive nano-electromechanical systems (NEMS)  
\cite{Yang:2001bf, Yang:2006cl, Li:2007ex} and simultaneously hosts highly-coherent spin centers, such as silicon vacancies ($\mathrm{V_{Si}}$) \cite{Riedel:2012jq} and divacancies ($\mathrm{VV}$) \cite{Falk:2013jq}.  Recently, mechanical tuning \cite{Falk:2014fh} and acoustic coherent control \cite{Whiteley:2019eu} of the $\mathrm{VV}$ spin-1 centers  in SiC has been experimentally demonstrated. 

In this work, we develop a theory of the spin-mechanical interaction for spin-3/2 qudits in SiC. Such centers can exist in a superposition of four states, which makes them promising for quantum computation and sensorics \cite{Soykal:2016tk, Soltamov:2018wb}. We determine the constant of spin-lattice interaction  from the temperature dependence of the spin-relaxation time, and use it to describe the interaction between spin qudit modes  and vibrational modes of a SiC membrane. We also discuss realistic applications of such a hybrid quantum system. Particularly, we propose an all-optical protocol for the DC magnetometry, where the sensitivity is defined by the mechanical Q-factor of the membrane. We also consider all-optical cooling of vibrational modes in a SiC membrane at room temperature, when they interact with a dense $\mathrm{V_{Si}}$ spin ensemble, and suggest an all-optical protocol for chemisorption measurements based on the mass-dependent shifts of the mechanical modes. Finally we discuss how the static strain of the membrane can be mapped via the shift of the zero-field splitting and propose to use it for force or acceleration measurements. 

\section{Spin-phonon interaction}

While our theoretical approach is general, we use the experimental parameters for  the so-called $\mathrm{V_{Si}}$(V2) spin qudit in 4H-SiC \cite{Sorman:2000ij} to link it to practical applications. It has spin $S = 3/2$ in the ground state \cite{Kraus:2013di}, which is split in two spin sublevels $m_S = \pm 3/2$ and $m_S = \pm 1/2$ with the zero-field splitting $ 2 D = 70 \, \mathrm{MHz}$ \cite{Kraus:2013vf}. First, we discuss how such spin centers interact with lattice vibrations.

Vibrations create local deformations that affect the spin states associated with point defects in the crystal. The Hamiltonian, describing this spin-phonon interaction, can be constructed using the group representation theory \cite{Soykal:2016tk, Udvarhelyi:2018bx, Udvarhelyi:2018tg}. For the sake of simplicity, we use the spherical approximation, where the interaction Hamiltonian reads 
\begin{align}
 H_\text{def} = \Xi u_{\alpha\beta} S_\alpha S_\beta \,. \label{Vdef}
\end{align}
Here, $\Xi$ is the deformation potential constant that quantifies the effect of local stress on the qudit fine structure, $u_{\alpha\beta}$ is the deformation tensor, $\bm S = (S_x,S_y,S_z)$ is the vector of spin-3/2 operators.  We note, that Hamiltonian \eqref{Vdef} should be  even in spin operators due to the time-inversion symmetry. The recent \textit{ab initio} calculations suggest  that the spin-phonon interaction in SiC can be anisotropic \cite{Udvarhelyi:2018tg}. While taking into account the real low symmetry of the defect may lead to a small correction to our results, the conclusions drawn remain qualitatively unchanged.

In case of static deformation,  Eq.~\eqref{Vdef} describes modification of the zero-field splitting and spin level mixing, as discussed in Sec.~\ref{sec:acc}.

When $u_{\alpha\beta}$ is regarded as a deformation induced by phonons passing by the spin center, Eq.~\eqref{Vdef} can be used to calculate the rate of the direct transitions $W^\text{(direct)}_{m_S',m_S}$ between the sublevels with the spin projection on the $c$-axis $m_S$ and $m_S'$~\cite{Abragam_Bleaney}. We get 
\begin{align}\label{Wd}
W^\text{(direct)}_{\pm 3/2,\pm 1/2} =  \frac{\Xi^2 |E_{\pm 3/2} -E_{\pm 1/2} |^2 }{2 \pi \hbar^4 \rho \bar v^5 }  k_B T \,,
\end{align}
where $E_{m_S}$ is the energy of the spin sublevel with $m_S = \pm 1/2, \pm 3/2$, $T$ is the temperature, $k_B$ is the Boltzmann constant, $\rho$ is the mass density, and $\bar v = 5v_l v_t/(2v_t+3v_l)$ is the averaged velocity of longitudinal and transverse phonons. The spin transition rate $W^\text{(direct)}_{\pm 3/2,\mp 1/2}$ is given by the same Eq.~(\ref{Wd}), where $E_{\pm 1/2}$ should be replaced with $E_{\mp 1/2}$. The other spin transition rates vanish, $W^\text{(direct)}_{+3/2,-3/2} = W^\text{(direct)}_{+1/2,-1/2} = 0$. The presence of spin transitions   with the spin projection change by $\pm 2$ as well as by $\pm 1$ is the direct consequence of the interaction Hamiltonian~\eqref{Vdef} being quadratic in spin operators.

With increasing temperature, the Raman processes that involve absorption of a thermal-energy phonon followed by its reemission start to give the dominant contribution to the spin relaxation~\cite{Abragam_Bleaney}. The corresponding transition rates read 
\begin{align}\label{WR}
W^\text{(R)}_{\pm 3/2,\pm 1/2} = W^\text{(R)}_{\pm 3/2,\mp 1/2}   = \frac{2\pi\Xi^4 k_B^5}{15\hbar^7 \rho^2 \bar v^{10}} T^5 \,,
\end{align}
and the rates $W^\text{(R)}_{+3/2,-3/2}$ and $W^\text{(R)}_{+1/2,-1/2}$ are twice higher. The temperature dependence of the spin-lattice relaxation time $T_1$ of the $\mathrm{V_{Si}}$(V2) spin qudit in 4H-SiC has been experimentally investigated in detail \cite{Simin:2017iw,Fischer:2018fj}. It has been observed that $1 / T_1$ increases linearly with temperature as $A_1 T$ up to $30 \, \mathrm{K}$ and follows $A_5 T^5$ at high temperatures, in accord with Eqs.~\eqref{Wd} and~\eqref{WR}. Using the parameters of 4H-SiC (table~\ref{SiC-parameters}) together with the experimental value $A_5 = 1.1 \times 10^{-9}\,{\rm s}^{-1}/{\rm K}^5$ \cite{Simin:2017iw}, we estimate the deformation potential constant $\Xi \approx 2\,$meV. It gives the direct transition rate  $A_1= 0.2 \times10^{-2}\,{\rm s}^{-1}/{\rm K} $ for the energy difference of  $|E_{\pm 3/2} -E_{\pm 1/2} | = 500\,\rm MHz$ in Eq.~\eqref{Wd}, which is within the same order of magnitude with the experimental value $A_1 = 1.0 \times 10^{-2}\,{\rm s}^{-1}/{\rm K}$ \cite{Simin:2017iw}. The small discrepancy may be related to the spread of the SiC parameters in the literature and presence of additional relaxation mechanisms.

 Importantly, the spin-phonon interaction is affected strongly by the structure design. In what follows we present theoretical results for spin centers coupled to quantized mechanical vibrations of a rectangular membrane. 

\begin{table}[t]
\caption{Mechanical parameters of 4H-SiC \cite{Landolt-Boernstein}}
\begin{center}
\begin{tabular}{r|l}
Mass density & $\rho =3.2 \,{\rm g}/{\rm cm}^3$ \\
\hline
Young modulus &   $E = 4.7 \times 10^{12}$\,dyn/cm$^2$ \\
\hline
Poisson ratio &  $\sigma = 0.14 $  \\
\hline
Velocity of &  \\
longitudinal phonons  & $ v_l = 12.2 \times 10^5\,{\rm cm}/{\rm s}$ \\
transverse phonons  & $ v_t = 8.0 \times 10^5\,{\rm cm}/{\rm s}$ 
\end{tabular}
\end{center}
\label{SiC-parameters}
\end{table}

\section{Quantized mechanical vibrations of a membrane}

\begin{figure}[t]
\includegraphics[width=.47\textwidth]{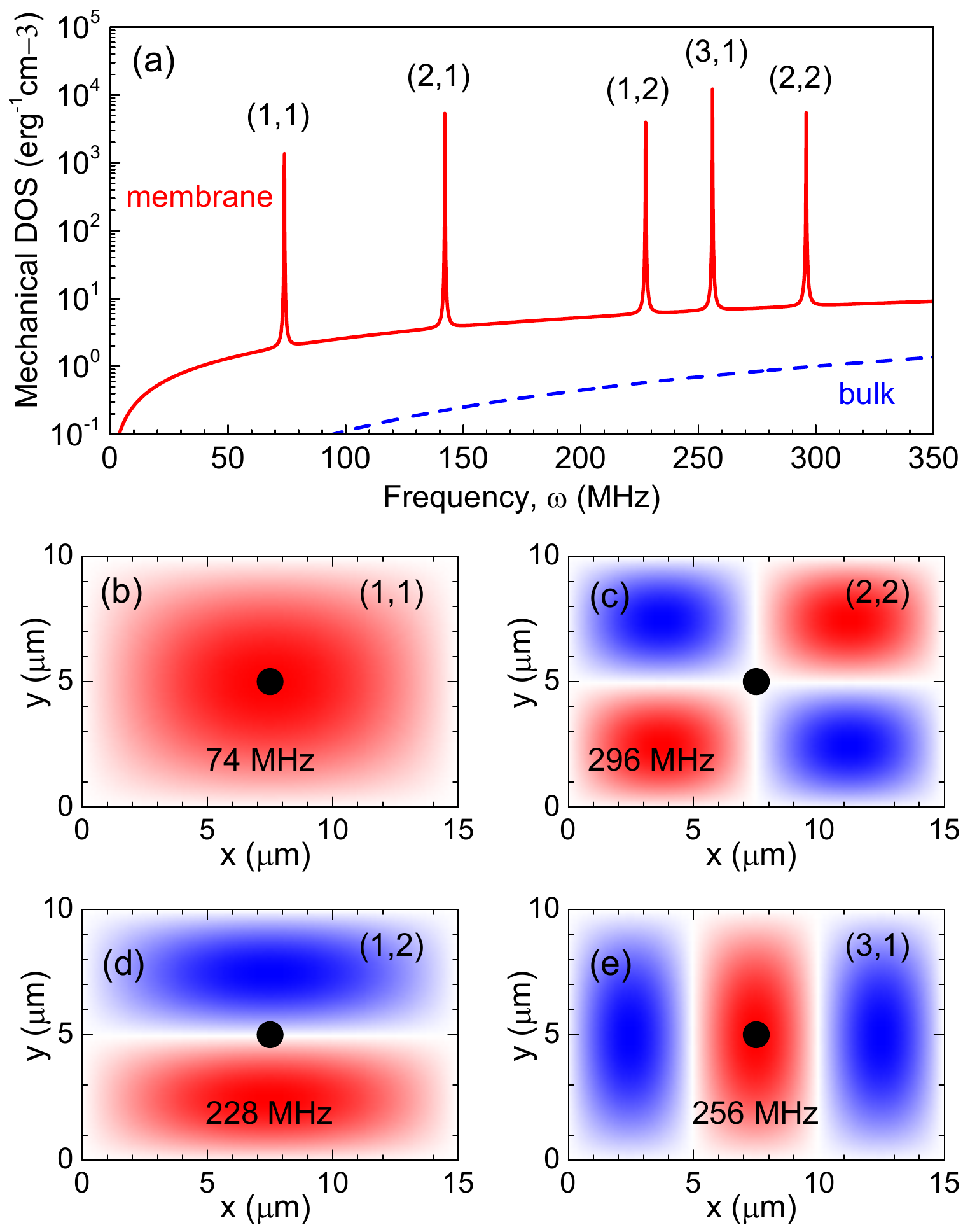}
\caption{Membrane vibrational eigenmodes. (a) Mechanical DOS in a $10\, \mu{\rm m} \times 15\, \mu{\rm m} $ SiC membrane with $Q = 10^4$ (solid line). The sharp peaks correspond to vibrational modes. For comparison, mechanical DOS in bulk SiC is shown by the dashed line. (b)--(e) Displacement distribution maps  for various vibrational modes (labeled in each plot). Red and blue color corresponds to positive and negative displacement along the membrane normal. The solid dot in the middle of the membrane indicates the position of the $\mathrm{V_{Si}}$ centers. }
\label{fig1}
\end{figure}

To be specific, we consider  a rectangular membrane with dimensions $L_x \times L_y = 10\, \mu{\rm m} \times 15\, \mu{\rm m} $ (Fig.~\ref{fig1}), the thickness $h=1\, \mu{\rm m}$, and suppose that initially it is not stressed. Particularly, such a membrane can be fabricated on the 4H-SiC platform using epitaxial growth in combination with dopant-selective photoelectrochemical etch, as has previously been used for the fabrication of high-Q  photonic crystals \cite{Bracher:2015gg}. We assume that the mechanical quality factor of the membrane is $Q = 10^4$  \cite{Barnes:2011gr, Villanueva:2014jc}.  

The system can support mechanical vibrations characterized by (i) in-plane and (ii) out-of-plane displacement. Both of them have only the in-plane component of the wave vector. The reason is that the out-of-plane wave vector component is quantized, which would result in frequencies $\sim \pi \bar v /h$, which are higher than the spin transition frequencies considered below.  The modes with in-plane displacement are similar to the acoustic waves in bulk material, with the same transverse sound velocity $v_t = \sqrt{E/[2\rho(1+\sigma)]}$ and a slightly different longitudinal sound velocity $v_l'=\sqrt{E/[\rho(1-\sigma^2)]}$~\cite{Landau:7}.   Here $E$ is the Young modulus $\sigma$ is the Poisson ratio of the material. 
   
The out-of-plane membrane displacement $\zeta$ is governed  by the equation
 \begin{align}
h \rho \frac{\partial^2 \zeta}{\partial t^2} + \mathcal{D} \Delta^2 \zeta = 0 \,,
 \end{align}
 where $\mathcal D =  Eh^3/[12(1-\sigma^2)]$ and $\Delta$ is a two-dimensional Laplace operator. An appropriate boundary conditions, determined by how the membrane is attached to the environment should be imposed. For the sake of simplicity, we assume that the edges are supported.  The eigenmodes of the system then read~\cite{Landau:7}
 \begin{align}
 \zeta^{(j)}(x,y) = \sqrt{\frac{2\hbar}{\rho h \omega_{j}L_xL_y}}\sin\left(\frac{\pi n_x x}{L_x}\right)\sin\left( \frac{\pi m_y y}{L_y} \right) , 
 \label{Modes}
 \end{align}
and the corresponding eigenfrequencies are 
 \begin{align}
 \omega_{j} = \sqrt{\frac{\mathcal{D}}{h\rho}}  \left[ \left(\frac{n_x}{L_x} \right)^2 +  \left(\frac{n_y}{L_y} \right)^2  \right] , 
 \label{Eigenfrequencies}
 \end{align}
where $j=(n_x,n_y)$ with integer $n_x,n_y  \geq 1$ enumerating the vibrational modes. 

The mechanical density of states (DOS) for the parameters of table~\ref{SiC-parameters} is presented in Fig.~\ref{fig1}(a). It  is calculated as
 \begin{align}
\bar{D}(\omega) = \frac{\omega(v_l'^{-2}+v_t^{-2})}{2\pi h}+ \frac{1}{\pi hL_xL_y}\sum_j \frac{\Gamma_j}{(\omega-\omega_j)^2+ \Gamma_j^2} \,,
 \end{align}
where the first term stems from the in-plane vibrations freely propagating in the membrane and the surrounding, while the second term represents the confined out-of plane modes, and $\Gamma_j= \omega_j/(2Q)$ is the decay rate of the vibrational mode $j$.   The displacement for several membrane eigenmodes, $j=(1,1)$, $(2,2)$, $(1,2)$ and $(3,1)$ is illustrated in Figs.~\ref{fig1}(b)-(e), respectively. 

For comparison, we also show in Fig.~\ref{fig1}(a) by dashed curve the DOS of 3D phonons in the bulk material given by $\bar{D}(\omega) = \omega^2(v_l^{-3}+2v_t^{-3})/(2\pi^2)$. The superior mechanical DOS of the membrane and the presence of mechanical resonances makes this system favorable for the study of spin-mechanical effects.

\section{Results and discussion}

When the frequency of a vibrational mode matches the energy difference between two spin sublevels, the rate of corresponding spin transitions is drastically enhanced. One the one hand, this results in a speed-up of the qudit spin relaxation time; on the other, leads to a deviation of the steady-state number of vibrational quanta in the mechanical mode from its equilibrium value. We discuss below optical protocols for detection of these effects and describe possible applications of such  optically detected spin-mechanical resonance (ODSMR). 

\subsection{Low-temperature ODSMR magnetometry}

\begin{figure}[t]
\includegraphics[width=.47\textwidth]{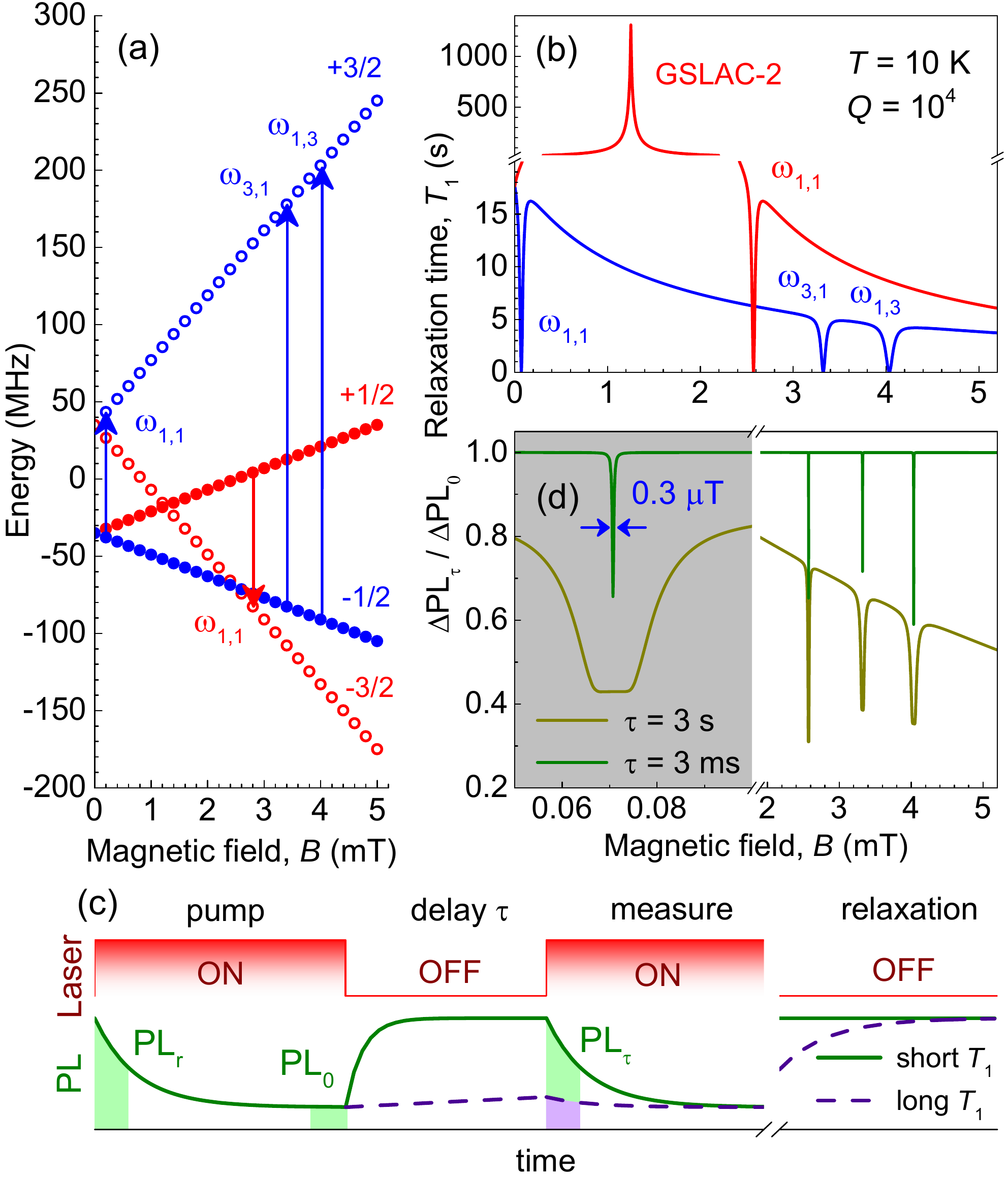}
\caption{Spin-mechanical magnetometry. (a) The fine structure of the  $\mathrm{V_{Si}}$(V2) ground state in the external magnetic field $B_z$. The solid (open) circles represent the preferentially populated (depleted) states under optical pumping. The vertical arrows indicate available spin transitions frequencies $\omega_j$ due to interaction with membrane modes $j = (n_x,n_y)$ in case when the $\mathrm{V_{Si}}$ qudit is placed in the middle of the membrane, as shown in Fig.~\ref{fig1}(b)--(e). (b) Spin relaxation time $T_1$ between the  $-1/2 \rightarrow +3/2$ (red) and $+1/2 \rightarrow -3/2$ (blue) spin sublevels. The deeps correspond to the membrane modes. The peak at $1.25 \, \mathrm{mT}$ is due to the ground state level anticrossing.  (c)  An all-optical protocol to probe spin-mechanical modes. 
The solid and dashed line correspond to the PL variation in case of short (resonant) and long (non-resonant) spin relaxation time. (d) Relative PL difference $\Delta \mathrm{PL}_{\tau} / \Delta \mathrm{PL}_{0}$ with $\Delta \mathrm{PL}_{0, \tau} = \mathrm{PL}_{r} - \mathrm{PL}_{0, \tau}$ as a function of $B_z$ for different delays $\tau$. The grey area highlights the optically detected spin-mechanical resonance at $70 \,\mathrm{ \mu T}$ with calculated FWHM of  $0.3 \,\mathrm{ \mu T}$ for $\tau = 1$\,ms. }
\label{fig2}
\end{figure}

The idea of the magnetometry is illustrated in Fig.~\ref{fig2}(a).  Upon application of the external magnetic field $B_z$ along the $c$-axis, the $\mathrm{V_{Si}}$(V2) spin states are split and shift linearly with $B_z$. Optical excitation results in the preferential population of the $m_S = \pm 1/2$ states [solid circles in Fig.~\ref{fig2}(a)] \cite{Riedel:2012jq}. After the excitation is switched off, the spin relaxation $ \pm 1/2 \rightarrow \pm 3/2$ occurs. At low temperature, this relaxation is caused by the absorption or emission of single phonons, and typical relaxation time is on the order of $10$~seconds  \cite{Fischer:2018fj}. However, in certain magnetic fields, when the splitting between some spin sublevels is equal to a membrane eigenfrequency [vertical arrows in Fig.~\ref{fig2}(a)], the spin relaxation is accelerated due to the resonantly enhanced probability of phonon emission or absorption.  

The spin transitions transitions can be described in terms of spin-mechanical interaction Hamiltonian Eq.~\eqref{Vdef},  where the oscillating deformation $u^{(j)}_{\alpha\beta}$ is induced by the periodical mechanical displacement of the membrane $ \zeta^{(j)}$, Eq.~\eqref{Modes}. 
In a thin membrane, the deformation is distributed linearly along the normal of the membrane and its maximal value
\begin{align}\label{umem}
u^{(j)}_{\alpha\beta} = -\frac{h}{2} \frac{\partial^2 \zeta^{(j)}}{\partial r_\alpha \partial r_\beta} \  (\alpha = x,y), \ \  u^{(j)}_{zz} = \frac{h}{2} \frac{\sigma}{1-\sigma} \Delta \zeta^{(j)} 
\end{align}
occurs at the membrane surfaces. It what follows, we assume that the spin centers are created close to the surface and in the center of the membrane, as shown schematically in Fig.~\ref{fig1}(b-e), unless explicitly mentioned. Technically, this can be realized using focused ion beam \cite{Kraus:2017cka}. 

We now calculate the spin transition rates induced by the interaction with the membrane vibrations.  The transitions with the spin projection change by $\Delta m_S = \pm 1$ are governed by $u_{xz}$ and $u_{yz}$ strain components, that vanish for the mechanical membrane modes. The spin transition rates with $\Delta m_S =\pm 2$  [vertical arrows in Fig.~\ref{fig2}(a)], caused by the vibrations in mode $j$, are given by
\begin{align}\label{Wvib}
W^{(j)}_{\pm 3/2, \mp 1/2} =  \frac{3 }{2}  \frac{\Xi^2 |u_{xx}^{(j)} \pm 2\rmi u_{xy}^{(j)} - u_{yy}^{(j)}|^2  \Gamma_j   N_j}{(|E_{\pm 3/2}-E_{\mp 1/2}|-\hbar\omega_{j})^2+\hbar^2\Gamma_{j}^2} \,,
\end{align}
where $N_j$ is the number of phonons in the mode, which is given  by $\overline N_j = k_B T/(\hbar\omega_j)$ in the thermal equilibrium.

Figure~\ref{fig2}(b) shows the spin relaxation times $(\sum_j W^{(j)}_{ + 3/2, -1/2})^{-1}$ (blue) and $(\sum_j W^{(j)}_{ - 3/2, +1/2})^{-1}$ (red), calculated as a function of the magnetic field using Eq.~\eqref{Wvib}. At certain magnetic fields, the spin relaxation time drops down by two orders of magnitude. This occurs when the spin splitting between the $m_S = +3/2$ and $m_S = -1/2$ states or between the $m_S = -3/2$ and $m_S = +1/2$ states is equal to the eigenfrequency $\omega_{j}$ of mode  $j=(n_x,n_y)$ with odd $n_x$ and $n_y$. Indeed, according to Eq.~\eqref{Modes}, only such ``bright'' modes have non-zero deformation in the middle of the membrane and hence interact with spin centers [Figs.~\ref{fig1}(b)-(e)]. Furthermore, our calculations indicate the possibility to increase of the relaxation time by two orders of magnitude ($> 1000 \, \mathrm{s}$) at $B_z = 1.25 \, \mathrm{mT}$ due to the second ground state level anticrossing (GSLAC-2)  \cite{Simin:2016cp}. 

Figure~\ref{fig2}(c) presents an all-optical protocol to detect the spin-mechanical modes. The system is excited by a sequence of two short optical pulses, leading to the preferential population of the $m_S = \pm 1/2$ states with respect to $m_S = \pm 3/2$ states.  The photoluminescence (PL) of $\mathrm{V_{Si}}$  is spin-dependent, i.e., contains a contribution proportional to the population difference between the  $m_S = \pm1 /2$ and $m_S = \pm 3/2$ states~\cite{Riedel:2012jq}. Therefore, the population difference induced by the first pump pulse can be deduced from $\Delta \mathrm{PL}_{0} = \mathrm{PL}_{r} - \mathrm{PL}_{0}$, where $\mathrm{PL}_{r}$ is the reference PL recorded immediately after the pump laser is switched on and $\mathrm{PL}_{0}$ is recorded at the end of the pump pulse, when the spin pumping has taken place. After the pump pulse, the  photo-induced population difference decays due to the spin relaxation processes  with the rates of Eq.~(\ref{Wvib}). To measure this decay, the system is excited with the second optical pulse after delay $\tau$. 
The difference between the PL intensity induced by the second pulse and the reference PL intensity  $\Delta \mathrm{PL}_{\tau} = \mathrm{PL}_{r} - \mathrm{PL}_{\tau}$ is measured.

Figure~\ref{fig2}(d) shows $\Delta \mathrm{PL}_{\tau} / \Delta \mathrm{PL}_{0}$ as a function of the magnetic field calculated using the spin relaxation time of Fig.~\ref{fig2}(b) for two values of delay $\tau$. As expected, the pronounced dips correspond to the resonances with ``bright'' mechanical modes of the membrane. The full width half maximum (FWHM) of these deeps depends on $\tau$, becoming narrow for shorter delay times. The grey area in Fig.~\ref{fig2}(d) zooms in on the ODSMR  at $70 \,\mathrm{ \mu T}$.  For $\tau = 1$\,ms, we obtain the FWHM  $\Lambda_\omega = 0.3 \,\mathrm{ \mu T}$.  For comparison, a typical FWHM of the optically detected magnetic resonance (ODMR) lines associated with the  $\mathrm{V_{Si}}$ centers in SiC is about $100 \,\mathrm{ \mu T}$ \cite{Kraus:2013di}. In case of ODSMR the magnetic field sensitivity depends on the mechanical quality factor $Q$ and spin relaxation time $T_1$, while in case of ODMR it is determined by the spin coherence time $T_2$. For $T_1 \gg T_2$ and high $Q \gg  \omega_{j} T_2 $, the ODSMR-based protocol provides higher sensitivity. Assuming 100\% efficiency of the spin read-out \cite{Baranov:2011ib, Nagy:2018ey} and pumping \cite{Fischer:2018fj}  at low temperature and the photon count rate from a single $\mathrm{V_{Si}}$ defect $R = 4 \times 10^4 \, \mathrm{Hz}$ \cite{Widmann:2014ve, Fuchs:2015ii}, we estimate DC magnetic field sensitivity $\delta B_\text{min} = \Lambda_\omega / \sqrt{R N_d} \sim 1/\sqrt{N_d} \,\, \mathrm{nT \cdot Hz^{-1/2}}$, where $N_d$ is the number of the $\mathrm{V_{Si}}$ defects. We note that for dense spin ensembles, inhomogeneous broadening can eliminate the advantage of the ODSMR protocol. 

\subsection{Room-temperature nonlinear effects in the strong  pumping regime}

In the above discussion, we assume that the number of quanta in the vibrational modes corresponds to the environment temperature. However, when a single mechanical mode interacts with many spin centers that are driven from thermal equilibrium by optical pump, the effective temperature of the mode can deviate from that of the environment \cite{MacQuarrie:2017hl}. In case of spin-3/2 centers, this process can be qualitatively understood from Fig.~\ref{fig2}(a). At a magnetic field of $ 2.57241 \, \mathrm{mT}$, corresponding to ODSMR with the $\omega_{1,1}$ mode, the spin relaxation leads to phonon emission. As a result, the number of phonons increases, which can be described as an increase of the effective temperature of the $\omega_{1,1}$ mode. On the contrary, at a magnetic field of $ 0.07072 \, \mathrm{mT}$, also corresponding to ODSMR with the $\omega_{1,1}$ mode, the spin relaxation leads to phonon absorption.  As a result, the effective temperature of the $\omega_{1,1}$ mode decreases.

To describe these heating and cooling processes under ODSMR, we use the rate equation approach.  
The dynamics of the spin level occupancies $f_{m_S}$ ($m_S=+3/2,+1/2,-1/2,-3/2$) in an ensemble of spin centers under optical pumping is given by the equation set
\begin{align}\label{eq:rateD}
\frac{df_{s}}{dt} &=  \sum_{m_S'} \Big( W^\text{(R)}_{m_S,m_S'}+ \sum_jW^{(j)}_{m_S,m_S'}  \Big) (f_{m_S'} -f_{m_S}) \nonumber\\
&+ I P_{m_S,m_S'} f_{m_S'}\,,
\end{align}
where, $I$ is the pump intensity, $P$ describes pump-induced transitions, and $W^{(j)}$ is given by Eq.~\eqref{Wvib}. Since the phonon, involved in the Raman process of spin relaxation, have high energy and do not feel the confinement,  we use the bulk expression of Eq.~\eqref{WR} for $W^\text{(R)}$. 
In order to describe the preferential population of  $\pm 1/2$ states under optical pumping, we
assume that, in the PL cycle, the spin center can go from the $\pm 3/2$ state to the $\pm 1/2$ state with the rate $\eta I$, and can come back  from the $\pm 1/2$ state to the $\pm 3/2$ state with the rate $\eta' I$.  The corresponding optical pump matrix reads 
\begin{align}
P = \left( \begin{array}{cccc} 
-\eta & \frac{\eta'}{2} & \frac{\eta'}{2}  & 0 \\
 \frac{\eta}{2} & -\eta' & 0 & \frac{\eta}{2}  \\
 \frac{\eta}{2} & 0 & -\eta' & \frac{\eta}{2}  \\
0 & \frac{\eta'}{2} & \frac{\eta'}{2}  & -\eta 
\end{array} \right) \,.
\end{align}
Then, the maximal spin polarization degree achieved at high intensities $I \to \infty$ is $\Delta f = f_{+3/2}+f_{-3/2}-f_{+1/2}-f_{-1/2} = (\eta-\eta')/(\eta+\eta') $.
The experimentally obtained value $f_{+1/2}-f_{+3/2} \approx f_{-1/2}-f_{-3/2}  \approx 0.03$ for the $V_{\mathrm{Si}}$ centers at room temperature \cite{Fischer:2018fj} gives $\eta/\eta' \approx 1.1$.  
 
 The transition rates $W^{(j)}$ are proportional to the numbers of the vibrational quanta $N_j$. In order to determine them,  we use the corresponding rate equations, 
 \begin{align}\label{eq:rateP}
\frac{dN_{j}}{dt} &=  \sum_{m_S',m_S} \theta(E_{m_S'}-E_{m_S}) N_d W^{(j)}_{m_S,m_S'}  (f_{m_S'} -f_{m_S})  \nonumber\\
& -  ( N_j - \overline N_j ) \frac{\omega_j}{Q},
\end{align}
 where $N_d$ is the number of spin centers that interact with mode $\omega_j$, and $\theta$ is the Heaviside function. The simultaneous solution of coupled non-linear equations \eqref{eq:rateD}-\eqref{eq:rateP} allows us to calculate the steady-state spin level occupancies $f_{m_S}$ and vibration quanta numbers $N_j$. The PL intensity change is then given by 
 \begin{align}
 \frac{\Delta\text{PL}}{\text{PL}}=  \alpha  \Delta f
 \label{D_PL}
 \end{align}
where the dimensionless parameter $\alpha \approx 0.1$ quantifies the efficiency of optical spin read-out at room temperature \cite{Tarasenko:2017ky, Fischer:2018fj}.

\begin{figure}[t]
\includegraphics[width=.47\textwidth]{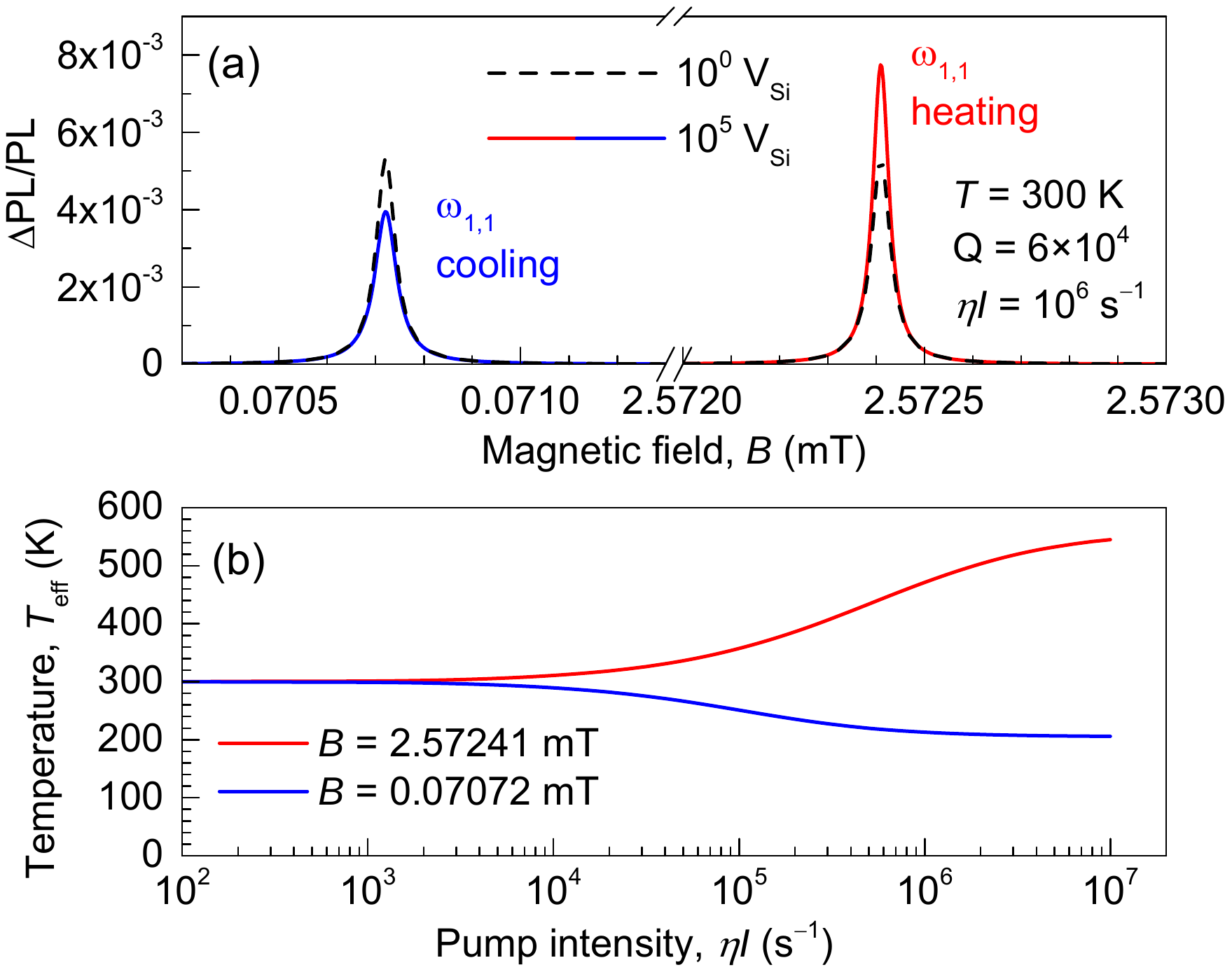}
\caption{Non-linear effects in the strong  pumping regime. (a) Relative change $\Delta \mathrm{PL} / \mathrm{PL}$ as a function of the magnetic field applied along the $c$-axis for pump intensity $\eta I = 10^6\,\text{s}^{-1}$. The peaks at $ 0.07072 \, \mathrm{mT}$  and $2.57241 \, \mathrm{mT}$ correspond to ODSMR with the vibrational mode (1,1). The dashed and solid lines are the calculation for a single $\mathrm{V_{Si}}$ and $10^5$ $\mathrm{V_{Si}}$ centers, respectively. The difference in the peak amplitude indicates heating and cooling of the vibrational mode.  (b)  Effective temperature of the  vibrational mode (1,1) as a function of pump intensity for $10^5$ $\mathrm{V_{Si}}$ centers at two magnetic fields corresponding to ODSMR. }
\label{fig3}
\end{figure}

The results of these calculations for mechanical quality factor $Q = 6 \times 10^{4}$ are presented in Fig.~\ref{fig3}(a). In case of a single $V_{\mathrm{Si}}$ center ($N_d = 1$), our model predicts sharp peaks in the PL intensity as a function of the magnetic field. The heights of these peaks are equal for the same mechanical mode. Particularly, the dashed line in Fig.~\ref{fig3}(a) shows the ODSMR lines for the lowest mode (1,1). 
For a large number of spin centers ($N_d = 10^5$) located close to the middle of the membrane, the nonlinear effects become visible, as shown by the solid lines in Fig.~\ref{fig3}(a).  Due to lower steady-state number of vibration quanta $N_j$ at $ 0.07072 \, \mathrm{mT}$  (phonon absorption) and higher at $2.57241 \, \mathrm{mT}$ (phonon emission), the ODSMR lines have different heights. 

The difference of  $N_j$ from its equilibrium value $\overline N_j$ can be interpreted as optically induced mode heating or cooling, with the effective temperature  of the mode given by $T^{(j)}_\text{eff} = N_j\hbar\omega_j / k_B$.  The effective temperature depends on the pump intensity $\eta I$ as shown in Fig.~\ref{fig3}(b). The heating/cooling processes can be quite efficient at high pump intensities. Remarkably, our model predicts optical cooling of the lowest vibrational mode (1,1) from room temperature to approximately $200 \, \mathrm{K}$ for  $\eta I  > 3 \times10^6 \, {\rm s}^{\mathrm{-1}}$. 

To estimate the laser powers required for the observation the aforementioned nonlinear effects, we first determine the characteristic pump intensity $\eta I_0$ required to induce spin polarization. To do this, we consider Eq.~(\ref{eq:rateD}) with $W^{(j)}_{m_S,m_S'} = 0$ (spin relaxation is dominated by the Raman process as in the bulk). Then,  in the steady-state, we have $\eta I_0 \sim W^\text{(R)}_{\pm 3/2,\pm 1/2} \sim 3 \times10^3 \, s^{\mathrm{-1}}$. This value is related to the characteristic laser power of about $0.1 \, \mathrm{mW \, \mu m^{-2}}$ \cite{Fischer:2018fj}. We assume that the $V_{\mathrm{Si}}$ ensemble with an area of $\mathrm{10 \, \mu {\rm m}^{2}}$ is created close to the middle of the membrane. With $N_d =10^5$ $V_{\mathrm{Si}}$ centers, this corresponds to the density of $10^{16} \, \mathrm{cm^{-3}}$. Putting all together, we conclude that the pump rate $\eta I$ of  $3 \times10^6 \, {\rm s}^{\mathrm{-1}}$, required to observe pronounced cooling effects of the membrane modes, can be achieved with a laser power of about  $1 \, \mathrm{W}$. From the absorption cross section of the $V_{\mathrm{Si}}$ centers \cite{Fuchs:2015ii, Fischer:2018fj}, we estimate that only about 5\% of $1 \, \mathrm{W}$ are absorbed in the $h = 1 \,  \mu {\rm m}$ thick membrane, minimizing heating effect due to the direct laser absorption.

 \subsection{Mass detection with ODSMR}
 
\begin{figure}[t]
\includegraphics[width=.47\textwidth]{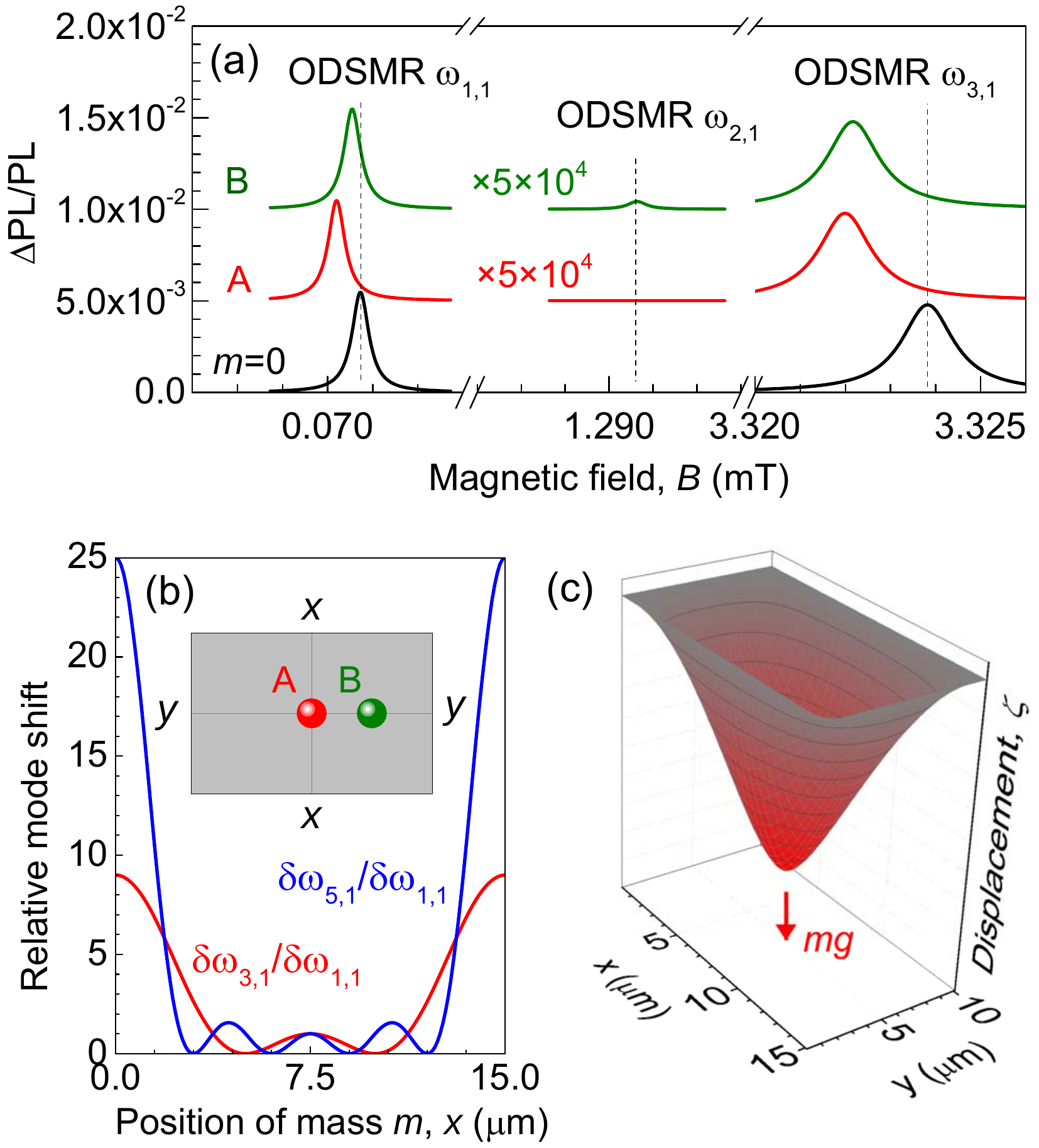}
\caption{Mass measurement of a probe particle attached to the membrane. (a) Relative change $\Delta \mathrm{PL} / \mathrm{PL}$ as a function of the magnetic field for positions A and B of the probe mass. The calculations are performed for $m / M = 0$ (the lowest curves) and $m / M = 10^{-4}$ with mechanical $Q$-factor $10^4$. The peaks at $ 0.0707 \, \mathrm{mT}$,  $1.2906 \, \mathrm{mT}$ and $3.3238 \, \mathrm{mT}$ correspond to ODSMR with vibrational modes (1,1), (2,1) and (3,1), respectively. (b) Relative difference of the mass-induced shift of the ODSMR lines $\delta\omega_{n_x,1} / \delta\omega_{1,1}$ (with $n_x = 3, 5$) as a function of the position of the probe mass $m$.  The inset indicates positions A and B on the membrane.  (c) Displacement of the membrane $ \zeta (x,y)$ caused by the inertia force $m g$ for a particle placed in the center.} 
\label{fig4}
\end{figure}

The proposed technique of ODSMR can be also used to detect a variation of the membrane mechanical properties. Suppose a particle with small mass $m$ gets attached to the membrane at coordinates $(x,y)$. As a consequence, the resonant frequencies of the mechanical modes are shifted by
\begin{align}
\Delta\omega_{n_x,n_y} = -\frac{4 m\omega_{n_x,n_y}}{M} \sin^2 \frac{n_x \pi x}{L_x} \sin^2\frac{n_y \pi y}{L_y} \,,
\label{ODSMR_mass}
\end{align}
where $M = \rho h L_xL_y$ is the membrane mass. Figure~\ref{fig4}(a)  shows an example for the ODSMR signal calculated for $Q = 10^4$,  $m/M = 10^{-4}$ and different mass positions A  and  B. The position A corresponds to the membrane center ($x = 7.5 \, \mathrm{\mu m}$,  $y = 7.5 \, \mathrm{\mu m}$) while the position B corresponds to a shift along the $x$ axis ($x = 11.25 \, \mathrm{\mu m}$,  $y = 7.5 \, \mathrm{\mu m}$), as shown schematically in the inset of Fig.~\ref{fig4}(b). 
 
Using the ODSMR shifts corresponding to different modes, one can determine both the position of the attached particle and its mass. In particular, it follows from Eq.~(\ref{ODSMR_mass}) that 
\begin{align}
\frac{\delta\omega_{n_x,1}}{\delta\omega_{1,1}} = \frac{\sin^2( n_x \pi x/L_x)}{\sin^2 (\pi x/L_x)} \,,
\label{ODSMR_mass_x}
\end{align}
where $\delta\omega_{n_x,n_y} =\Delta\omega_{n_x,n_y}/\omega_{n_x,n_y} $.  The $x$-coordinate can be unambiguously found from Eq.~\eqref{ODSMR_mass_x} by analysing the shift of several $(n_x,1)$ modes with odd $n_x$ [Fig.\ref{fig4}(b)], and the $y$-coordinate can be found in the similar way by analyzing $(1,n_y)$ modes. The attached particle mass can be obtained as, e.g.,
\begin{align}
m = \frac{4 M  \delta\omega_{1,1}}{(3-\sqrt{\delta\omega_{1,3}/\delta\omega_{1,1}})(3-\sqrt{\delta\omega_{3,1}/\delta\omega_{1,1}})} \,.
\end{align}

The sensitivity of mass measurement is determined by the membrane mass $M$ and the measurement accuracy of the relative ODSMR shift $\delta\omega_{1,1}$. The latter depends on the mechanical quality factor $Q$ and the accuracy of $\mathrm{\Delta PL / PL}$ measurement,  Eq.~(\ref{D_PL}).  We estimate the minimum detectable  ODSMR shift as a product 
\begin{align}
\delta\omega_\text{min} = \frac{1}{Q_S} \frac{1}{\alpha \Delta f} \frac{1}{\sqrt{R N_d}} \,.
\label{ODSMR_min}
\end{align}
Here, $Q_S = \omega_j / \Lambda_\omega$ is an effective ODSMR Q-factor. We obtain from Fig.~\ref{fig4}(a) $Q_S = 10^2$. Note that for the protocol of Fig.~\ref{fig2}(c), $Q_S$ tends to the mechanical quality factor $Q$. The readout efficiency $\alpha$ and spin polarization $\Delta f$ yield $\alpha \Delta f = 3 \times 10^{\mathrm{-3}}$ at room temperature and  $\alpha \Delta f \sim 1$ at low temperature \cite{Baranov:2011ib, Nagy:2018ey, Fischer:2018fj}. The photon count rate from a single $\mathrm{V_{Si}}$ defect is $R = 4 \times 10^4 \, \mathrm{Hz}$ \cite{Widmann:2014ve, Fuchs:2015ii}. We then estimate $\delta\omega_\text{min} \sim 10^{-2} / \sqrt{N_d} \,  \mathrm{Hz^{-1/2}}$ at room temperature and $\delta\omega_\text{min} \sim 10^{-6} / \sqrt{N_d} \, \mathrm{Hz^{-1/2}}$ for the optimized protocol at cryogenic temperature. With the membrane mass $M = 0.5 \, \mathrm{ng}$ for the given dimensions, we obtain the mass sensitivity $\delta m_\text{min}  \sim 10  / \sqrt{N_d}  \,\, \mathrm{pg \cdot Hz^{-1/2}}$ and $\delta m_\text{min}  \sim 1  / \sqrt{N_d}  \,\, \mathrm{fg \cdot Hz^{-1/2}}$, respectively. According to these estimations, ultrahigh $\mathrm{zg}$ sensitivity can be achieved by increasing the mechanical (ODSMR) quality factor $Q$ ($Q_S$) and reducing the membrane mass $M$. This should allow mass detection of individual macromolecules~\cite{Yang:2006cl}.

Apart from the shift of the mechanical frequencies, the attached particle can also leads the mixing of vibrational modes.  
When a particle is placed not in the geometrical center of the membrane, the dark modes that in the unperturbed case are not detectable due to their symmetry, e.g., mode $(2,1)$, get an admixture of bright modes and emerge in the ODSMR spectra. 
Then, the relaxation rate in the vicinity of the dark mode frequency $\omega_j$ is given by 
\begin{align}\label{Wvib-1}
W^{(j)}_{\pm 3/2, \mp 1/2} =  \frac{3 }{2}  \frac{\Xi^2 |\sum_{j'}\alpha_{jj'} (u_{xx}^{(j')} \pm 2\rmi u_{xy}^{(j')} - u_{yy}^{(j')})|^2  \Gamma_j   N_j}{(|E_{\pm 3/2}-E_{\mp 1/2}|-\hbar\omega_{j})^2+\hbar^2\Gamma_{j}^2} \,,
\end{align} 
where $\alpha_{jj'} $ is the admixture amplitude of the mode $j'=(n_x',n_y')$ to the mode $j=(n_x,n_y)$ that reads
\begin{align}
\alpha_{jj'} = \frac{4m}{M} \frac{\omega_j^2}{\omega_{j'}^2-\omega_j^2} \,\sin\frac{n_x \pi x}{L_x}\sin\frac{n_y \pi y}{L_y}\sin\frac{n_x' \pi x}{L_x}\sin
\frac{ n_y' \pi y}{L_y} \,.
\end{align}
Figure~\ref{fig4}(a) shows the ODSMR  in the vicinity of  frequency of the ``dark'' mode $(2,1)$ calculated for $m / M = 10^{-4}$. The weak resonance is present only when particle is in position B, i.e., shifted from the membrane center. The amplitudes of the ``dark''-mode resonances can be used to determine the position of the particle and its mass, however this method is less sensitive compared to the method discussed above.

\subsection{ODSMR accelerometry}\label{sec:acc}

Finally, we discuss how the static membrane deflection induced by some external force can be detected.   
The material deformation caused by membrane deflection modifies the fine structure parameters of the spin center. To describe this, we rewrite the spin-mechanical interaction Hamiltonian~\eqref{Vdef} as
\begin{align}
H_\text{def} = \delta D  \left(S_z^2-\frac54 \right) + \frac12\left(\delta E\, S_+^2 + \delta E^* S_-^2 \right) \,,
\end{align}
where $S_\pm = S_x \pm S_y$, $\delta D = \Xi(u_{zz} - \frac12 u_{xx} - \frac12 u_{yy})$ describes variation of the zero-field splitting, and $\delta E = \Xi(\frac12 u_{xx} - \frac12 u_{yy}-\rmi u_{xy})$ mixes the $m_S = \pm 1/2$ and $m_S = \mp 3/2$ spin states. Using the expression for the deformation tensor components Eq.~\eqref{umem}, we obtain
\begin{align}
\delta D = \Xi \frac{h}{4} \, \frac{1+\sigma}{1-\sigma} \Delta \zeta \,.
\end{align}
The value of $\delta D$ is directly available as a shift of the ODSMR line. Measurement of this shift with spatial resolution can be used to map the membrane bending $\Delta \zeta (x,y)$.  

A possible application of the proposed bending measurement is a detection of the acceleration $g$. To this end, a particle of a large known mass $m $  is placed in the center of the membrane. The inertia force $m g$ produces a load in the membrane center. Figure~\ref{fig4}(c) presents membrane deflection $ \zeta (x,y)$ as a function of the $x$ and $y$ coordinates induced by such a load. The maximum membrane bending  is achieved in the membrane center and reads $\Delta \zeta_0 =  C  mg/\mathcal{D}$, where  $C$ is a constant of the order of unity, determined by the membrane and particle dimensions.  For the particle size $d \ll L_x,L_y$ and $m \gg M$, it is given by $C \approx (1/2\pi) \ln (d/L_x)$. Then, by measuring the zero-field splitting variation $\delta D$ one calculates the inertia force from
\begin{align}
\frac{m g}{\delta D} =  \frac{Eh^2}{3C(1+\sigma)^2\Xi}  \,.
\label{accel}
\end{align}
For the parameters  from table~\ref{SiC-parameters}, the deformation potential constant $\Xi \approx 2\,$meV, and the membrane thickness  $h=1\, \mu{\rm m}$, we estimate  from  Eq.~(\ref{accel}) that $m g / \delta D \sim  0.1 \, \mathrm{g \cdot cm \cdot s^{-2} / MHz}$. 

The minimum detectable zero-field splitting variation can be estimated using Eq.~\eqref{ODSMR_min} 
 as $\delta D_\text{min} =  D \delta\omega_\text{min}$.  We assume that  a gold ($19.3 \, \mathrm{g / cm^3}$) sphere with a diameter of $1 \,\mathrm{\mu m}$ is attached to the center of the membrane, which has a mass of $80 \,\mathrm{pg}$. Combining Eq.~\eqref{accel} and Eq.~\eqref{ODSMR_min}, we obtain the room-temperature acceleration sensitivity $g_\text{min} \sim 5\times 10^{8} / \sqrt{N_d}  \,\, \mathrm{cm \cdot s^{-2} \cdot Hz^{-1/2}}$.   In fact, this sensitivity is miserable because of membrane dimensions are not optimal for the acceleration measurement. 
The sensitivity can be dramatically improved by using very thing membranes. For instance, if the thickness of the membrane is $h = 10 \, \mathrm{nm}$ instead of $h = 1 \, \mathrm{\mu m}$, the sensitivity is improved by 4 orders of magnitude. Attaching a heavier particle to the membrane center is another way to improve the acceleration sensitivity. For instance, a gold sphere with a diameter of $10 \, \mathrm{\mu m}$ should give further improvement by 3 orders of magnitude. Thus, the room temperature sensitivity better than $g_\text{min} \sim 50 / \sqrt{N_d}  \,\, \mathrm{cm \cdot s^{-2} \cdot Hz^{-1/2}}$ is feasible though technologically challenging. Further improvement is possible using membranes with larger lateral dimensions, however the mechanical resonance frequencies shift to the sub-MHz range.

\section{Conclusions}

We have considered theoretically the optically detected spin-mechanical resonance associated with half-integer spin centers in SiC membranes. It is caused by the spin-phonon coupling and occurs when the conditions for the spin resonance  and mechanical resonance are simultaneously fulfilled. Due to the optical spin pumping mechanism and spin-dependent recombination, ODSMR  can be detected as  a change in the PL intensity. Based on these properties, we have proposed all-optical sensing protocols, where the sensitivity is determined  by the mechanical quality factor.  We have discussed the realistic conditions, under which the femtotesla-scale magnetic sensing and zeptogram-scale mass sensing can be achieved. By placing a micron-size particle at the center of the membrane, the ODSMR can be used as an accelerometer. In addition, we have considered the application of strong optical pumping of a dense spin ensemble in a membrane  for cooling of the mechanical modes from room temperature to below 200~K. Our findings suggest that hybride SiC spin-mechanical systems are a promising platform for quantum sensing applications.

\begin{acknowledgments}
This work has been supported by the German Research Foundation (DFG) under Grant  AS 310/5. A.V.P. also acknowledges the support by the Foundation for the Advancement of Theoretical Physics and Mathematics ``BASIS.'' 
\end{acknowledgments}


\end{document}